# On Phenomenology of Complex Scientific Systems

*Vladimir Shiltsev*

*Fermi National Accelerator Laboratory, PO Box 500, Batavia, IL 60510, USA*

*Abstract*

Performance evolution of a number of complex scientific and technical systems demonstrate exponential progress with time $\propto e^{+t/C}$. The speed of progress $C$ - a measure of difficulty and complexity – is analyzed for high energy elementary particle colliders, astrophysical searches for galaxies and exoplanets, protein structure determination and compared with computers and thermonuclear fusion reactors. An explanation of the characteristic exponential progress is offered.

Complex systems are commonly understood as a highly structured, hierarchical systems, with large number of independent interacting components, with multiple evolution pathways and usually those difficult to understand and predict – see, e.g., Ref.[1] for many examples including genetic algorithms, computers, geophysical landscapes, the brain, the immune systems, protein folding, the stock market, etc. At the same time, there is no agreed-upon definition of the complexity of the real-life systems, because mathematical constructs – like Kolmogorov's algorithmic complexity [2] – are rarely easily applicable. In this Letter we study of the complexity of scientific problems on the base the difficulty of solving them. We consider a number of remarkable large systems associated with several fundamental problems and show that in the past they exhibited exponential growth of their performance $\propto e^{+t/C}$ over significant time intervals. The extracted speed of the progress $C$ gives a quantitative measure of complexity, which one can intuitively agree with.

*Particle Accelerators*

Particle accelerators are an excellent example of complex scientific systems – since 1920's, they were widely used for understanding the nature of nuclear energy and discoveries of many elementary particles and fundamental laws of high energy physics. Below we consider in detail a proton-antiproton collider Tevatron at FNAL (Batavia, IL, USA), which was in operation for two and a half decades as the world highest energy accelerator before conceding to the Large Hadron Collider at CERN in 2010. The unique measure of performance of any collider is luminosity *L*, that defines how many particle reactions of interest *N* are generates per unit time *dN/dt* :

$$dN/dt = L \cdot \sigma_r \qquad (1)$$

where $\sigma_r$ is the reaction cross-section. Mathematically, the luminosity is a product of several factors such as frequency of collisions, number of particles in each of the colliding beams, and inverse cross-section of the beams overlap [3]. Technically, several accelerators are needed to prepare the required beams – for example, in the case of Tevatron, these are Linac, Booster, Main Injector, Recycler, Debuncher, Antiproton Accumulator and the Tevatron ring itself [4]. Each of the accelerators in turn requires a number of technical subsystems which have to work perfectly in order for the entire complex to be effective in producing the reactions. Such systems include magnets (e.g., the Tevatron employs almost 800 state-of-the-art superconducting magnets), ultra-high vacuum system, radiofrequency acceleration, beam collimation, particle detectors, antiproton production targets and beam-lines, beam cooling systems, beam stabilization systems, beam diagnostics, control system, cooling water, personnel safety, high voltage and high current elements and power supplies. Optimization of the luminosity factors requires solution of a number important beam physics issues such as antiproton production, storage and cooling, beam-beam effects, transverse and longitudinal beam instabilities, space-charge effects in low energy beams, halo formation and losses. Some 500 peoples including almost 100 PhD physicists take part in operating the Tevatron accelerators. Altogether, the system of the Tevatron collider is quite complex, it has at least three levels of structural hierarchy (elements, individual accelerators, complex of machines) augmented by interconnections of various effects.

The Tevatron luminosity history in 2001-2011, during so called Collider Run II period [4] is presented is Fig.1. Each point represent a maximum peak luminosity achieved in a month of operation. Overall, one can see that the performance increased gradually and the progress was due to numerous improvements, some of which were implemented during operation, and others introduced during regular annual machine shutdown periods. Detail analysis [5] indicated that as many as 30 improvements addressing all the parameters affecting the luminosity resulted in a 50-fold increase of luminosity from $L_i \approx 8\times10^{30}$ cm$^{-2}$s$^{-1}$ to $L_f \approx 400\times10^{30}$ cm$^{-2}$s$^{-1}$, or about 14% per step on average (varying from varying from few % to some 40% with respect to previously achieved performance level). In general, the complex percentages, i.e. "*N*% gain per step, step after step, with regular periodicity" explain the *exponential* growth of the luminosity

$$L(t_0+T)=L(t_0)\times e^{T/C} \quad (2).$$

The pace of the luminosity progress was not always constant. As one can see from Fig.1, the Collider Run II luminosity progress was quite fast with $C\approx0.7$ year in the period from 2001 to mid-2002 of the complex startup; stayed on a steady exponential increase path with $C\approx2.0$ yr from 2002 till 2007, and significantly slowed down afterward, $C\approx8.6$. Other high energy particle colliders show very similar features of the luminosity evolution (see Fig.2): usually, the very fast progress during the start-up period is followed by extended period of time with exponential growth of the performance which fades when the all the possibilities and ideas for further improvements are fully explored and luminosity stabilizes at its ultimate level. Table I summarizes the coefficients *C* for various colliding facilities.

The evolution of the performance of continuously improving facilities where every next step brings *x-fold* improvement on top of previous improvement can be further simplified in an approximate formulae:

$$C \cdot P = T \quad (3)$$

where the factor *P=ln(luminosity)* is the "*performance*" gain over time interval *T*, and *C* is a machine dependent coefficient equal to average time needed to increase the luminosity by *e=2.71…* times, or boost the "performance" *P* by 1 unit. Both, *T* and *C* have dimension of time, and the coefficient *C* can called and has the meaning of the "*complexity*" of the machine, as it directly indicates how hard or how easy was/is it to push the performance of the individual machine. In general, one can rightfully guess that the complexity *C* should be dependent on how well understood are the physics and technology of the machine, type of particles, efforts and resources invested into operation and upgrades of the system, number of elements and subsystems. For example, if a system *S* consists of a number of subsystems

$$S = S_1 \otimes S_2 \otimes S_3 \otimes ... \qquad (4)$$

then, its performance progress Eq.(2) is determined by complexities of its parts:

$$\frac{1}{C_S} \leq \sum_i \frac{1}{C_{S_i}} \qquad (5).$$

For example, if one breaks the Tevatron luminosity in three key factors such as (1) the number of protons, (2) number of antiprotons and (3) the geometrical beam compression factor – see Fig. 3 – then the breakout of the complexity $C=2.0$ is $C_1 \approx 20$, $C_2 \approx 2.8$, $C_3 \approx 6.2$.

Interestingly, not only luminosity but the energy of the particle accelerators, exhibits the *CPT*-like progress Eq.(3) with $C \approx 4.3$ for proton machines and $C \approx 5.2$ for electron accelerators. This fact has been known since long ago, often represented in the form of so called "Livingston plot" and explained as the result of evolution of acceleration techniques and instruments each consequently being built to exceed the energy of the predecessor by a factor of 2-7 [6]. Another example of the exponential progress with $C \approx 7.2$ is given in Fig.5 which presents the record proton pulse intensities achieved in various types of particle accelerators [7].

*Other Complex Scientific Systems*

The exponential growth is characteristic to advances in other areas of science and technology. Over the past decades, sky surveys have proven the power of large data sets for answering fundamental astrophysical questions. While photographic surveys of 20-th century covered large area, the data were not as usable as digital data and did not go as faint. Since 1980's new types of surveys employing CCD cameras allow to scan the sky about 100 times faster. Figure 6 from [8] charts the number of galaxies discovered by digital optical sky surveys over the past 25 years. It exhibits a clear exponential *CPT*-like growth with $C=3.0$ (straight line). This observational progress was based on a synergy of advances in telescope construction, detectors, and information technology and has had a dramatic impact on nearly all fields of astronomy, and areas of fundamental physics. Over approximately the same period of time another branch of observational astronomy - search for extrasolar planets - has progress exponentially as well: from initial discoveries to some 100 planets detected every year. Fig.7 summarizes the data from [9] together with a straight line corresponding to $C=4.2$.

Almost four decades of research on determination of protein structures have generated a wealth of data. Modern large-scale structural genomics facilities can determine the structures of a hundred or more proteins per year, with unprecedented high quality, providing a foundation for understanding macromolecules whose biological roles are known now and for those whose roles will be identified in the future. Technology development has played a critical role in structural genomics and rapid deposition of data in public databases has increased the impact and usefulness of the data [10]. Fig.8 plots the rate of annual depositions of protein structures in the Protein Data Bank – the central point of accumulation of the protein information worldwide [11]. Again, the progress is exponential, with the complexity coefficient $C=4.2$ in 1975-2005, while relative slow down – a period of increased complexity in the sense of Eq.(3) - starting afterward.

Fusion power is another example of extremely difficult and complex scientific problem. Authors of Ref.[12] noted :"…Our understanding of nuclear fusion and of nuclear fission emerged in the 1930's. Although fission reactors started delivering power during the following decade, it's only 6 decades later that a modest 16MW of fusion power were produced for a second by the JET (Joint European Torus) tokamak sited at Culham in the UK. Why is fusion power generation so much more difficult?" The answer is multifaceted mix of physics reasons, technology challenges – like the development of the materials necessary to withstand the extreme conditions inside a commercial reactor, needed depth of the understanding of various issues, and (limited) available resources – all that makes the fusion very complex. Fig. 9 from [12] depicts four decades of progress toward achieving a self-sustaining thermonuclear reaction in a magnetically confined plasma. The key figures of merit is so called "the fusion triple product" of the ion temperature, density and confinement time. It has to reach about $7\times10^{27}$ degree m$^{-3}$ s in the International Thermonuclear Experimental Reactor, and so far it had the *CPT*-like exponential growth with $C=2.4$ (straight line). Ultra-high power lasers are being considered as an alternative way to ignite the fusion, and their impressive progress in 1975-2000 [13] can be approximated by $C=3.3$ - see Fig.10.

The most cited example of the exponential growth of performance of a complex system is the "Moore's Law" [14] that describes about half a century trend in the history of computing hardware, namely, that the number of transistors that can be placed on an integrated circuit(IC) has doubled approximately every two years, yielding $C=2/\ln(2)=2.9$. It is of interest to note, that the density the elements on the IC is just one of the contributors to the pace for faster computers. There are many other advances in the field (architecture, communication means, clock speed, etc) which led to significantly more impressive progress of performance of the world's fastest computers – see Fig.11 – with $C=1.6$ [15]. Similarly, a steady improvement over the years of light-emitting diodes – LEDs – is summarized by "Haitz's Law" [16]. It states that every decade, the amount of light generated per LED package increases by a factor of 20, and the cost per lumen (unit of useful light emitted) falls by a factor of 10, for a given wavelength (color) of light. That law corresponds to $C=3.3$ (see Fig.11).

*Discussion and conclusions*

All the examples of complex systems considered above are summarized in Table II together with their calculate complexities. Errors in the values of $C$ are for standard r.m.s. deviations from the best exponential fit. One can see that exponential *CPT*-like performance progress is typical for many scientific and technical systems – the fact noted by many and reflected in various empirical "laws" similar to the ones considered above – e.g., "Kryder's law" (that magnetic disk areal storage density doubles annually), "Nielsen's law" (network connection speeds for high-end home users would increase 50% per year, or double every 21 months), "Rock's law" (the cost of a semiconductor chip fabrication plant doubles every four years), "Butter's law" of photonics (the amount of data coming out of an optical fiber is doubling every nine months), "Wirth's/Gate's/Page's law" (observation that the speed of commercial software generally slows by fifty percent every 18 months thereby negating all the benefits of the "Moore's law") [17]. The underlying explanation for the exponential performance progress is the fact that in many systems the improvements come in steps, and the goal for each step is set as percentage (*m*-percent) increase or *x*-fold increase with respect to what is already achieved, so after *n* steps, the performance is either $(1+m/100)^n \approx e^{nm/100}$ or $x^n = e^{n \ln(x)}$. From the comparative Table II, one can argue that increasing the energy of particle accelerators was significantly more complex problem (in the sense of difficulty and pace of the performance progress) than, say, improvement of the speed of computers. The suggested definition of the complexity, the coefficient $C$ in the exponential performance growth - see Eq.(2-3) – is well applicable to many large scientific and technical systems. Such a complexity factor reflects not only the scientific side of the problem, but also the social one - how important the problem is for the society, human and financial resources invested in its solution (e.g., the funding for the computer development exceeds the support of the exoplanet search by several orders of magnitude), etc. At the same time, it significantly differs from objectively defined complexity in mathematics. First of all, in many systems considered above, the complexity *C* varies with time, e.g. the progress slows down as soon as the system is well understood and the scientific or technical teams run out of ideas or lack of resources prevents further significant improvements. Another difference is that the apparent complexities of some systems are smaller than complexity of their parts – see Eq.(5). That holds for systems where the performance is a product of several factors (the luminosity of colliders, the fusion triple product, the speed of computers) – so, improvement of each factor helps the whole. For mathematical objects, the Kolmogorov complexity (the length of a shortest binary program to compute the process on a universal computer) of a system or a process is larger than complexity of any of its sub-programs or sub-processes.

TABLE I: "Complexities" of colliding beam facilities.

|  | *C* | *years* |
|---|---|---|
| SLC    *e+e-* | *1.6 ±0.1* | 1989-1997 |
| Tevatron Run II *p-pbar* | *2.0 ±0.2* | 2002-2007 |
| RHIC  *p-p* | *2.2 ±0.3* | 2000-2004 |
| HERA  *p-e* | *2.8 ±0.4* | 1992-00-2005 |
| SppS   *p-pbar* | *3.3 ±0.2* | 1982-1990 |
| LEP    *e+e-* | *3.3 ±0.3* | 1989-1995 |
| ISR    *p-p* | *3.7 ±0.3* | 1972-1982 |
| CESR  *e+e-* | *4.4 ±0.4* | 1984-1997 |

TABLE II: Progress rates ("complexities") of scientific and technical systems.

|  | *C* | *years* | *comment* |
|---|---|---|---|
| Fastest Computers | *1.6 ±0.1* | 1993-2010 | http://www.top500.org/ |
| Luminosity of Colliders | *1.6 …4.4* | 1972-2010 | see Table I |
| Fusion Reactors | *2.4 ±0.2* | 1969-1999 |  |
| Transistors per IC | *2.7 ±0.05* | 1971-2009 | Moore's Law |
| Galaxies Surveyed | *3.0 ±0.1* | 1985-1990 |  |
| Light per LED | *3.3 ±0.1* | 1969-2000 | Heitz's Law |
| Most powerful lasers | *3.3 ±0.5* | 1975-2000 | http://laserstars.org/ |
| Protein Structures | *4.2 ±0.2* | 1976-2010 | http://www.pdb.org/ |
| Exoplanets Search | *4.2 ±0.3* | 1991-2010 | NASA |
| Energy of accelerators | *5.2 ±0.3* | 1930-1990 | Livingston plot |
| Protons accelerated | *7.2 ±0.6* | 1960-2009 |  |

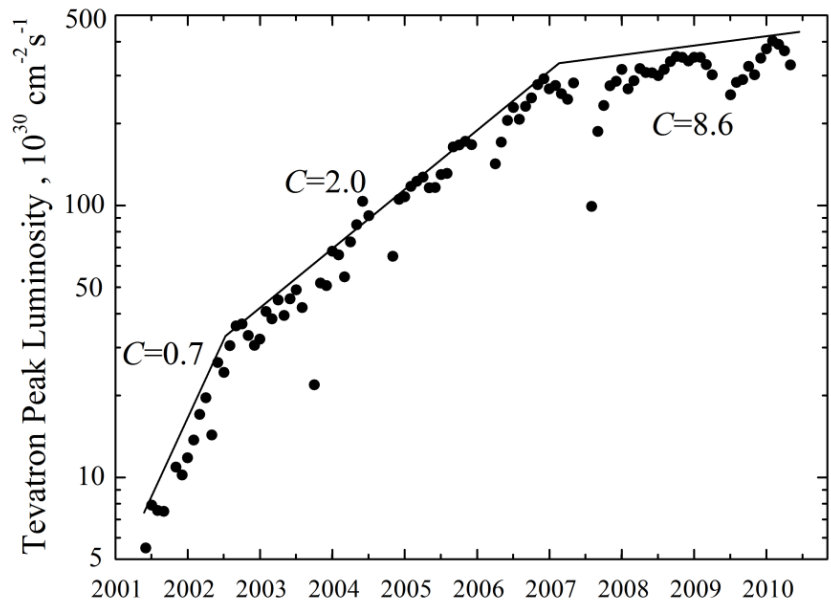

FIG.1: Tevatron peak luminosity progress during Collider Run II (2001-2011).

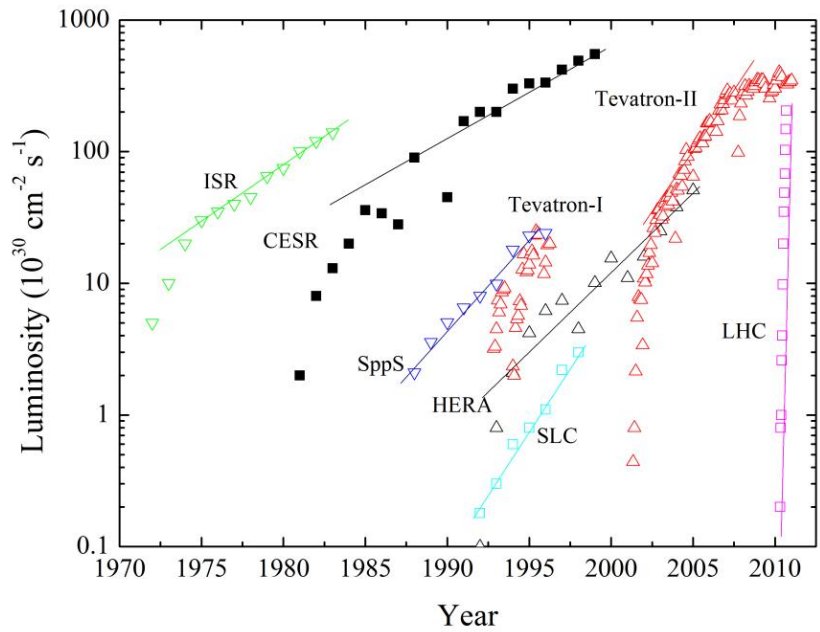

FIG.2: Luminosity of high energy particle colliders.

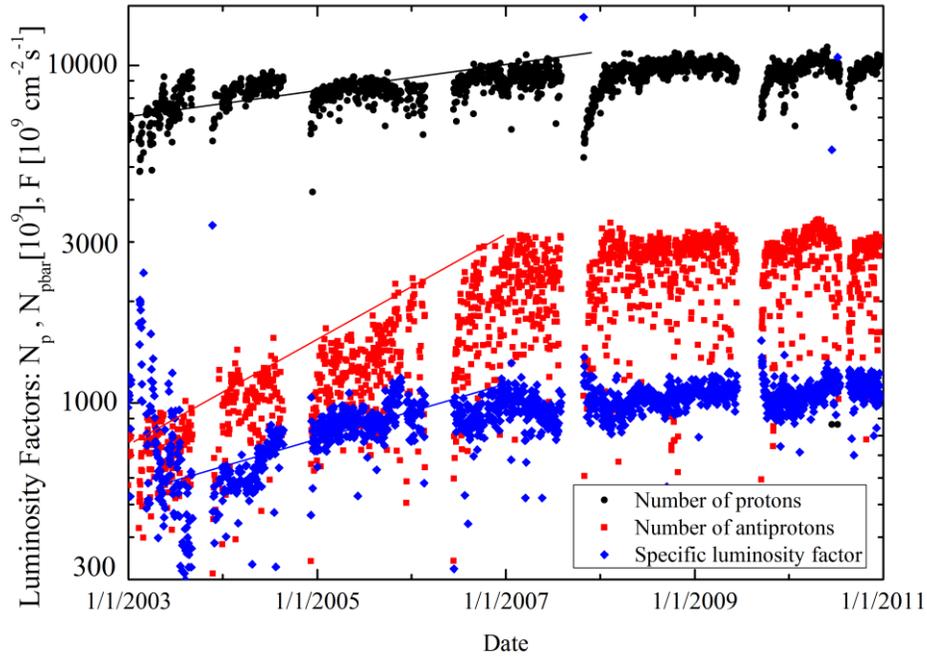

FIG.3: Progress on the Tevatron luminosity constituents: number of protons, number of antiprotons and beam compression factor.

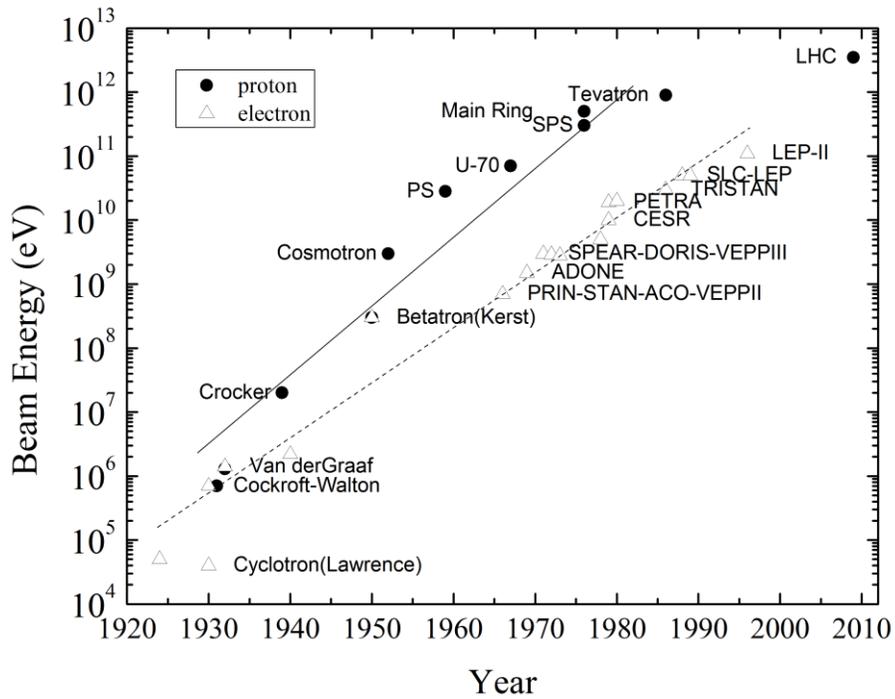

FIG.4: Highest energy particle accelerators (triangles − electron, circles − proton).

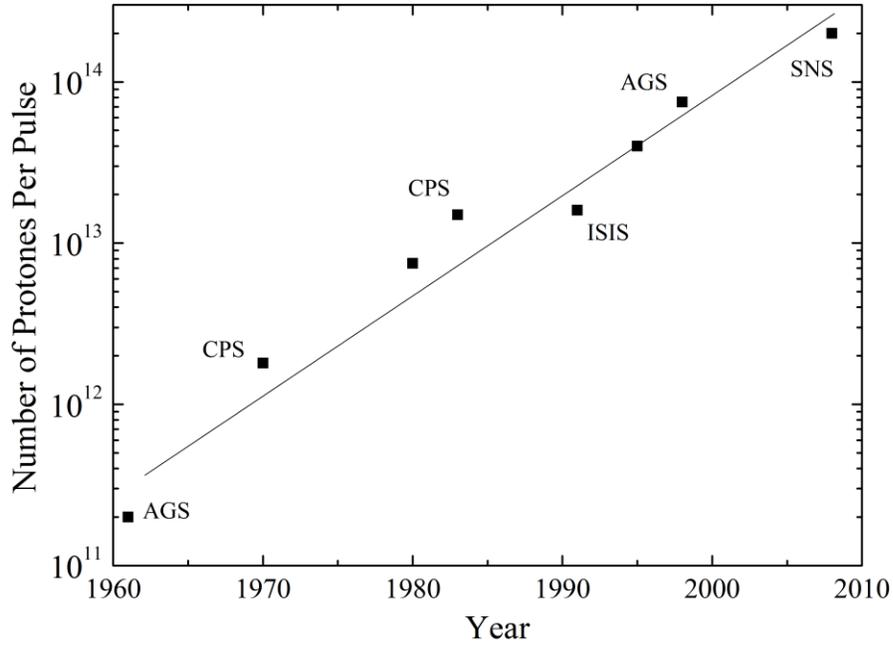

FIG.5: Evolution of proton beam intensity in accelerators (adapted from [7]).

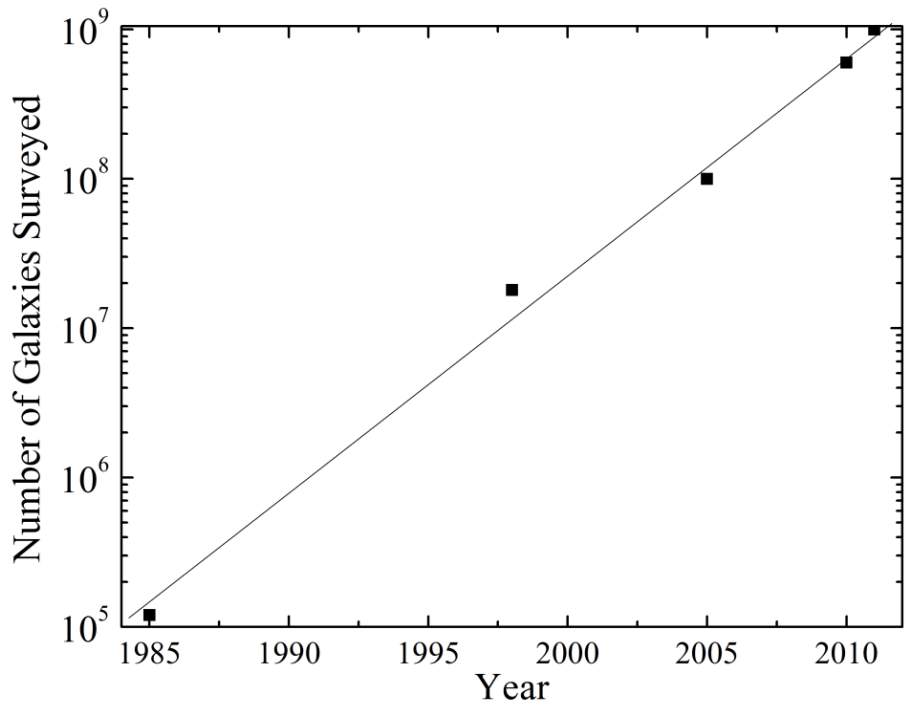

FIG.6: Number of galaxies surveyed by digital CCD telescopes (from [8]).

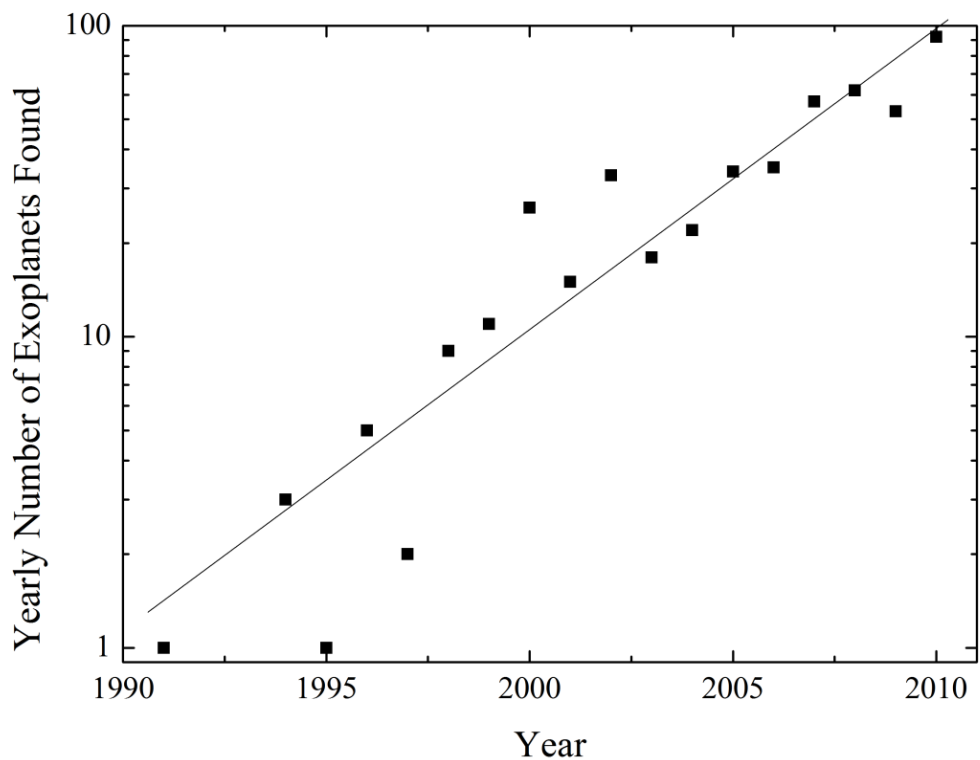

FIG.7: Yearly number of discovered extra-solar planets.

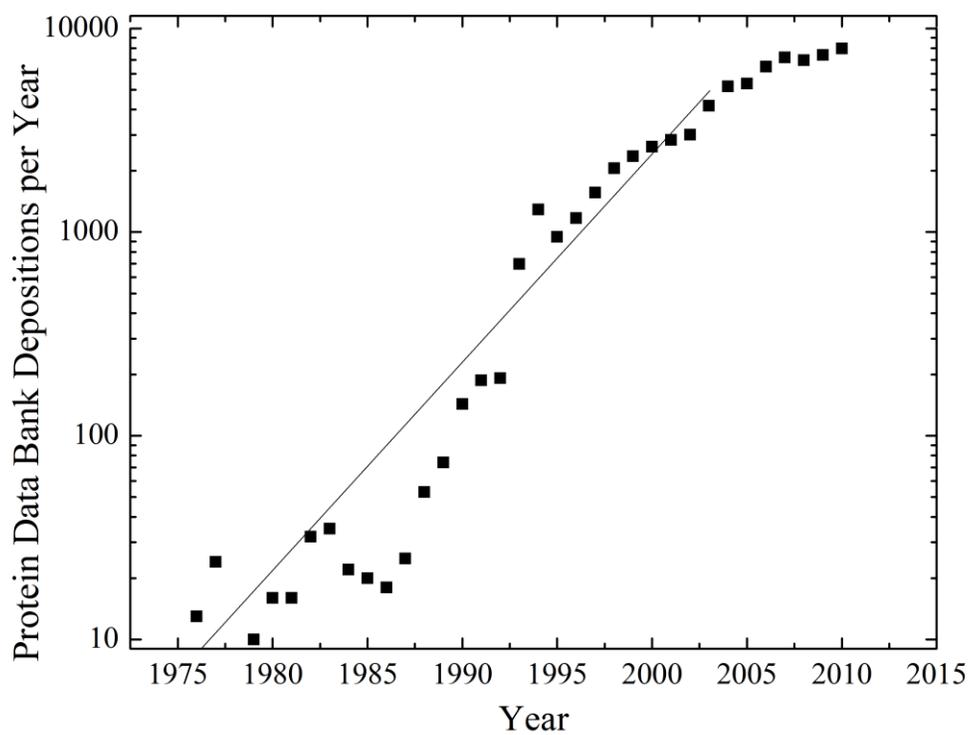

FIG.8: Number of protein structures deposited annually to the Protein Data Bank.

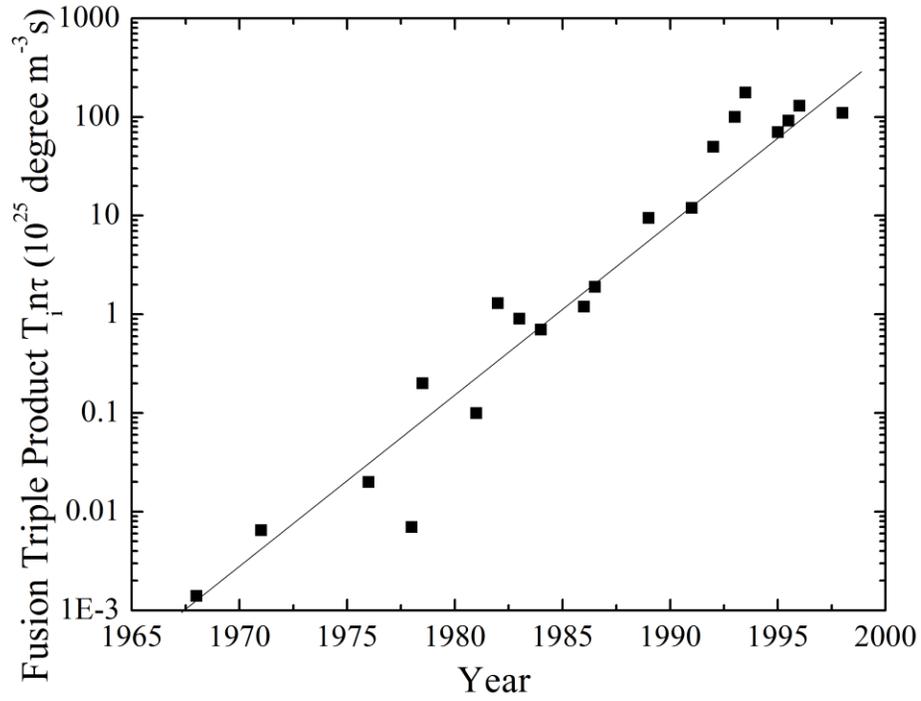

FIG.9: The fusion "triple product" of thermonuclear reactors (adapted with permission from Ref.[12]).

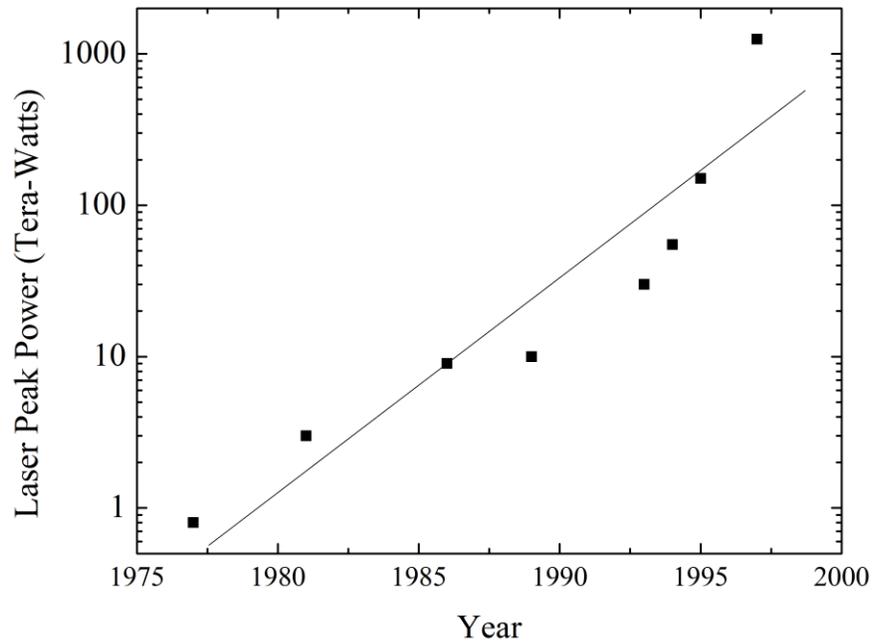

FIG.10: Evolution of peak power of the world's highest intensity lasers in 1975-2000.

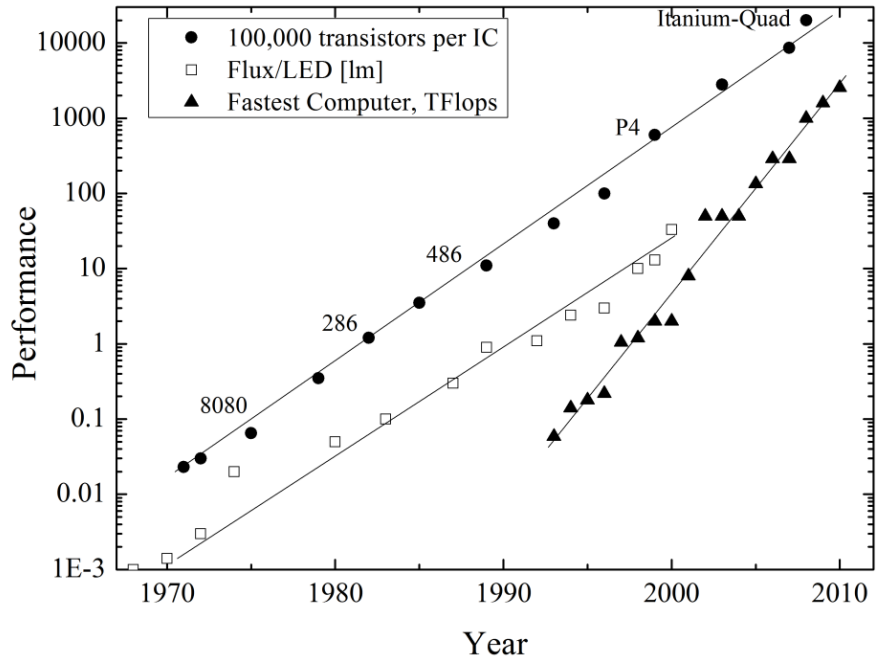

FIG.11: CPU transistor counts in 1971-2008 ("Moore's Law" – solid circles), amount of light generated per LED package ("Haitz's Law" - squares), and performance of the world's fastest computers in 1992-2010 (solid triangles).